\documentstyle[astrop-bib,proceedings]{crckapb}

\newcommand{\beq}{\begin{equation}}
\newcommand{\eeq}{\end{equation}}
\newcommand{\beqnn}{\begin{displaymath}}	
\newcommand{\eeqnn}{\end{displaymath}}		
\newcommand{\beqa}{\begin{eqnarray}}
\newcommand{\eeqa}{\end{eqnarray}}
\newcommand{\beqann}{\begin{eqnarray*}}
\newcommand{\eeqann}{\end{eqnarray*}}

\newcommand{\ben}{\begin{enumerate}}
\newcommand{\een}{\end{enumerate}}
\newcommand{\bit}{\begin{itemize}}
\newcommand{\eit}{\end{itemize}}
\newcommand{\bc}{\begin{center}}
\newcommand{\ec}{\end{center}}

\newcommand{\figref}[1]{Figure~\ref{#1}}

\newcommand{\eqref}[1]{Equation~\ref{#1}}

\newcommand{\Figref}[1]{\figref{#1}}

\newcommand{\eBoo}{\mbox{$\eta$~Boo}}

\def\Msol{\mbox{${M}_\odot$}}

\def\la{\mathrel{\hbox{\rlap{\hbox{\lower4pt\hbox{$\sim$}}}\hbox{$<$}}}}
\def\ga{\mathrel{\hbox{\rlap{\hbox{\lower4pt\hbox{$\sim$}}}\hbox{$>$}}}}

\def\laeq{\lower.5ex\hbox{{$\:\scriptstyle\buildrel < \over \sim\:$}}}
\def\gaeq{\lower.5ex\hbox{{$\:\scriptstyle\buildrel > \over \sim\:$}}}

\makeatletter 
 \def\sub#1{\relax\ifmmode _{\fam\z@ #1}\else
         $_{\fam\z@ #1}$\fi}
 \def\super#1{\relax\ifmmode ^{\fam\z@ #1}\else
         $^{\fam\z@ #1}$\fi}
\makeatother

\newcommand{\down}[2]{#1\sub{#2}}

\newcommand{\mycaption}[3]
{\if*#2 \caption{#3\label{#1}}
 \else  \caption[#2]{#3\label{#1}}
 \fi}

\newcommand{\comment}[1]{\relax}

\long\def\COMMENT#1\ENDCOMMENT{}
\def\ENDCOMMENT{}

\newcommand{\aCenA}{\mbox{$\alpha$~Cen~A}}

\newcommand{\Teff}{\down{T}{eff}}
\input{psfig}

\begin{opening}

\title{Observing solar-like oscillations}

\author{Hans Kjeldsen}
\institute{Teoretisk Astrofysik Center, Danmarks Grundforskningsfond,
Aarhus University, DK-8000 Aarhus C, Denmark}
\author{Timothy R. Bedding}
\institute{School of Physics, University of Sydney 2006, Australia}

\end{opening}
\begin{document}

\begin{abstract}
We review techniques for measuring stellar oscillations in solar-type
stars.  Despite great efforts, no unambiguous detections have been made.  A
new method, based on monitoring the equivalent widths of strong lines,
shows promise but is yet to be confirmed.  We also discuss several
subtleties, such as the need to correct for CCD non-linearities and the
importance of data weighting.
\end{abstract}

\vspace*{-1ex}
\section{Why search for solar-like oscillations?}
\vspace{-1ex}

Given the tiny amplitudes of oscillations in the Sun and the obvious
problems in detecting similar oscillations in other stars, we should first
ask whether the effort is justified.  Oscillation frequencies give
information about the sound speed in different parts of the stellar
interior.  They can be measured much more precisely than can any of the
other fundamental parameters which have been discussed at this meeting.
Accuracies of $10^{-3}$--$10^{-4}$ have been achieved for ``classical''
multi-periodic pulsators stars such as $\delta$~Scuti stars, rapidly
oscillating Ap (roAp) stars and $\beta$~Cephei stars.  These stars pulsate
with amplitudes typically 1000 times greater than seen in the Sun, so why
are we not satisfied with observing them?

One reason is that the classical pulsating stars are only found in
restricted areas of the HR diagram (the instability strips).  Since
oscillations in the Sun are thought to be excited by convective turbulence
near the surface, all stars with an outer convective zone should undergo
similar oscillations.  This makes it possible, at least in principle, to
perform seismic studies on {\em all\/} stars with spectral type later than
about~F5.

\begin{figure}
\centerline{\psfig{figure=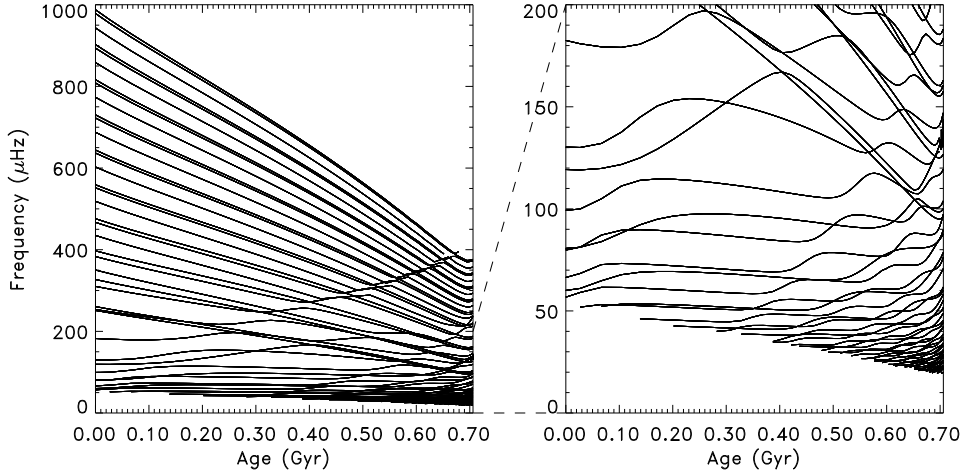,width=\the\hsize}}

\caption[]{Evolution of oscillation frequencies in a 2.2\,\Msol{} star,
from model calculations by J. Christensen-Dalsgaard.  Only modes with $\ell
= 0, 1, 2$ and $n\le 10$ are shown.  }
\label{fig.modes}

\vspace*{-1ex}
\end{figure}
\psfull

A second reason for studying solar-like oscillations is that the modes are
easy to identify.  There is little point in knowing the frequency of an
oscillation mode unless you also know in which part of the star that mode
is trapped.  An oscillation mode is characterized by three integers:
$n$~(the radial order), $\ell$~(the angular degree) and $m$~(the azimuthal
order)\footnote{In a star with no rotation or magnetic field, frequencies
do not depend on~$m$.}.  These specify the shape of the eigenfunction,
which in turn determines the sensitivity of the oscillation frequency to
the internal structure of the star.

\Figref{fig.modes} shows the oscillation frequencies of a non-rotating star
(mass 2.2\,\Msol) as it evolves.  At any instant during the star's
evolution, a vertical cross-section through this figure shows the
frequencies of oscillation modes with $\ell=0$, 1 and~2 (which are most
easily observed in an unresolved star).  However, in multi-periodic
$\delta$~Scuti and $\beta$~Cephei stars, only the lowest frequency modes
are found to be excited to an observable level, presumably due to the
details of the excitation process (the so-called $\kappa$~mechanism).  We
are therefore forced to identify modes in the crowded lower region of the
diagram.  To further complicate matters, these stars tend to be rapid
rotators, which causes a splitting of frequencies (analogous to Zeeman
splitting).  Finally, a given star is only observed to oscillate in a
seemingly random subset of possible modes.  Until reliable mode
identification is achieved, it will be impossible to apply asteroseismology
to these ``classical'' pulsating stars.

In contrast, it is easy to identify the modes of solar-like oscillations.
At least in the Sun, all modes in a broad frequency range are excited.
Furthermore, these modes approximately satisfy an asymptotic relation, with
modes of fixed $\ell$ and differing $n$ having regularly spaced frequencies
separated by the so-called large separation,~$\Delta\nu$.  The resulting
comb-like structure is clearly seen in the upper part of \figref{fig.modes}
and allows modes to be identified directly from the oscillation spectrum.

Measuring $\Delta\nu$ provides an estimate of the stellar density.
Moreover, the small differences between observed frequencies and those
predicted by the asymptotic relation give crucial information about the
sound speed deep inside the star.


\section{Sensitivities of detection methods}


\vspace{-1ex}
\paragraph{Velocity} 

In the Sun, the strongest modes have velocity amplitudes of about 25\,cm/s,
which corresponds to a wavelength variation ($\delta\lambda/\lambda$) of
less than $10^{-9}$, or 4.2\,$\mu$\AA{} at 5000\,\AA\@.  Detecting such
miniscule Doppler shifts in other stars is extremely difficult.
Spectrographs cannot be made with absolute stabilities of $10^{-9}$, so one
must simultaneously monitor the wavelength of a stable reference (e.g., a
Na or K resonance cell, an $I_{2}$ absorption cell or telluric absorption
features).  The noise levels at present are down to about 0.5\,m/s, which
is a factor of two higher than the solar signal.

\vspace{-2ex}
\paragraph{Radius} 

Given that solar periods are around 5 min, the change in radius is only
about 12\,m or 17 microarcseconds.  Astrometry of the solar limb using
SoHO/MDI has recently revealed the oscillations (J.~Kuhn et al., Proc.\
IAU Symp.\ 181, in press), but such observations will surely be impossible
for other solar-like stars.

\vspace{-2ex}
\paragraph{Intensity} 

The solar oscillations have been observed as variations in total intensity,
with amplitudes of about 4~ppm (parts per million).  Open clusters are a
natural target for differential CCD photometry and the lowest noise level
so far achieved is 5--7 ppm, from observations by \citeone{GBK93} of twelve
stars in M\,67 using six telescopes (2.5\,m to 5\,m) during one week.  This
is an interesting noise level, less than a factor of two away from solar
photometric amplitude.

Ground-based photometric observations are severely hampered by atmospheric
scintillation.  Several space missions have been proposed, but only one has
so far been launched: the EVRIS experiment, on board the Russian Mars96
probe, which ended in the Pacific Ocean.

\vspace{-2ex}
\paragraph{Temperature} 

Since the change in radius during solar oscillations is insignificant, the
intensity fluctuations observed in the Sun must result from local
temperature changes in the atmosphere of about 6\,mK ($\delta\Teff/\Teff
\approx 10^{-6}$).  It has been suggested that these temperature changes
can be measured by their effect on spectral absorption lines
\cite{KBV95,BKR96}.  For example, the Balmer lines in the Sun should show
variations in equivalent width of about 6\,ppm.  As discussed below, the
equivalent-width method has so far attained noise levels in other stars of
2--3 times the solar peak amplitude.

\section{Some subtleties}
\vspace{-1ex}

Achieving low noise levels demands care during both observing and data
analysis.  One major requirement is high efficiency, in order to get as
many photons as possible (photon counting statistics are a fundamental
limitation).  This requires optical systems with high transmissions,
detectors with high Q.E. (i.e., CCDs) and observations with a high duty
cycle.  This may force one to observe under quite unusual conditions.  For
example, in the case of photometry these requirements mean observing
defocussed stars in order to avoid saturating the CCD\@.

Linearity of the system is another important issue.  Measuring oscillations
at the ppm level requires that the detector be linear to the level of
$10^{-3}$ or better.  This is certainly not trivial and our tests of
different CCDs and controllers often reveal deviations from linearity of up
to a few per cent.  Unless correction is made for these effects, the extra
noise will destroy any possibility of detecting oscillations.

Each step in the data reduction procedure must be tested to establish how
much noise it adds to the time series.  It also helps if, as well as
measuring the parameter which is expected to contain the oscillation signal
(magnitude, velocity or the line strength), one also monitor extra
parameters.  For example, by correlating measured magnitudes with seeing
variations, one has a chance to remove the influence of seeing simply by
subtracting that part of the signal which correlates with seeing.  Of
course, this assumes that the real oscillations do not correlate with the
seeing.  This process of decorrelation, which can be repeated for other
parameters (total light level, position on detector, etc.), is very
powerful but can also be quite dangerous if not done with care.

Once a time series has been extracted, the search for oscillation
frequencies is done by calculating the power spectrum.  The simplest method
is to Fourier transform the time series and take the squared modulus.  The
resulting spectrum shows power as a function of frequency, and a
significant peak in this spectrum implies a periodic signal in the time
series data.  However, the standard Fourier transform treats all data
points as having equal weight.  In reality, data quality can vary
significantly within a data set, due to variable weather conditions or even
because data are being combined from different telescopes.  The power
spectrum is very sensitive to bad data points -- the final noise level will
be dominated by the noisiest parts of the time series.  One should
therefore calculate a weighted power spectrum, with each data point being
allocated a statistical weight according to its quality \citeeg{FJK95}.
Unfortunately, this procedure is not widely used and many published power
spectra have higher noise than necessary.

\section{Recent reults}
\vspace{-1ex}

Attempts to detect solar-like oscillations have been reviewed by
\citeone{B+G94} and \citeone{K+B95}, and here we only discuss more recent
results.  Most efforts have concentrated on subgiants, since these are
expected to have higher oscillations amplitudes than the Sun.

{\em \eBoo\/} is the brightest G-type subgiant.  We observed this star over
six nights with the 2.5-m Nordic Optical Telescope \cite{KBV95}.  Using the
equivalent-width method, we claimed a detection of solar-like oscillations
with amplitudes at the expected level and frequencies that were
subsequently shown to be consistent with models \cite{ChDBK95,Gu+D96}.
However, a search for velocity oscillations in \eBoo{} by \citeone{BKK97}
has failed to detect a signal, setting limits level below the value
expected on the basis of the \citename{KBV95} result.

Some support for the equivalent-width method was given by \citeone{KHB97},
who detected the 5-minute oscillations in the Sun from measurements of
H-beta equivalent widths.  However, they have subsequently had difficulties
in reproducing these results (Keller, priv.\ comm.).

{\em \aCenA\/} is the brightest G-type main-sequence star.  We obtained
H$\alpha$ spectra over six nights in April 1995 using the 3.9-m AAT (UCLES)
and the 3.6-m ESO (CASPEC).  Data reduction using the equivalent-width
method was hampered by a variability of the continuum, which seems to be
due to some kind of colour term in scintillation at a level of about
$10^{-4}$ per minute (well below the normal photometric scintillation).

{\em Procyon\/} is the brightest F-type subgiant.  Recent results from
Doppler-shift measurements are: (i)~\citeone{BCC95}, using a narrow-band
filter, have retracted an earlier possible detection; and
(ii)~\citeone{BKN96}, using an echelle spectrograph, have not detected a
signal.  We observed Procyon for several hours per night during the 1995
run mentioned above.  Preliminary analysis reveals excess power at the
expected amplitude and frequency, but sparse sampling prevents a definite
measurement of the frequency splitting.  A recent campaign on Procyon in
Jan--Feb 1997 by several members of SONG (see below) should produce results
soon.

{\em Arcturus\/} and similar red giants are variable in both velocity
(e.g., \citebare{H+C96} and references within; \citebare{Mer96}) and
intensity \citeeg{E+G96}, but the presence of solar-like oscillations has
not yet been established.

\section{Conclusion}
\vspace{-1ex}

In the last few years, the precision in velocity and photometric
measurements has not been significantly improved.  The new equivalent-width
method is far from being fully developed and no confirmation of the claimed
signal in \eBoo{} has been made.  Hopefully, the formation of SONG (Stellar
Oscillations Network Group; see {\tt http://www.noao.edu/noao/song/}),
which aims to do joint research in this field, will soon produce robust
detections of oscillation signals.


Space would be a wonderful place to do photometry.  Although COROT has been
selected, for now we will have to continue using ground-based facilities.
It is important to remember that we are only about a factor of two from
producing noise levels equal to the solar oscillation signal, and that some
stars are expected to oscillate with higher amplitudes than our own Sun.  A
network of 10-m class telescopes should provide scintillation levels low
enough for detection of oscillations in M\,67 \cite{GBK93}, but getting a
week on each of these big telescopes will not be easy.

We still await real asteroseismic results for solar-type stars.  However,
twenty-five years ago we were in a similar situation concerning
oscillations in the Sun.  First, people had to believe that these
oscillations actually existed.  Next, they had to measure their frequencies
accurately.  Finally, we have reached a stage where we truly see the Sun as
a physics laboratory.  The same will one day be true for other stars.  It
might take longer than twenty-five years, but it could also happen much
faster.

\medskip
\small

\noindent{\bf Acknowledgements}~~~This work was supported in part by the
Danish National Research Foundation through its establishment of the
Theoretical Astrophysics Center.  TRB is grateful for funding from the
University of Sydney Research Grants Scheme and the Australian Research
Council.

\vspace{-2.5ex}
\small


\normalsize

\section*{Discussion}
\tt\raggedright

ROBERT KURUCZ: I have a philosophical comment; do not take it personally.
We do not know the spectrum of a single star.  We would learn a lot more if
you would spend some of your observing time taking a high-resolution, high
signal-to-noise spectrum.

\medskip

HANS KJELDSEN: I agree that we would learn a lot from such a spectrum.
Firstly, however, it would not take very much time to do this for Procyon.
In fact, at $R\approx \tt200\,000$ you can reach S/N of 1000 in 10\,s and S/N
of 10\,000 in 15 minutes on a 4-m telescope.  Secondly, I disagree that
such a spectrum would provide more information than would detections of
$p$~modes.  The solar oscillations provide a huge amount of information and
tell us almost all that we know about the Sun.  This will never be the case
for a high S/N spectrum.

\medskip

J\O{}RGEN CHRISTENSEN-DALSGAARD (also responding to Kurucz): Observations of
the stellar radiative spectrum, however precise, would not give information
about the structure of the deep stellar interior.  One might perhaps be
forgiven for regarding such information as being more fundamental than fine
details of the surface properties and composition.

\end{document}